\documentclass[namedreferences]{solarphysics}

\usepackage[hyperref,optionalrh,showbiblabels]{spr-sola-addons} 
\usepackage{graphicx}        
\usepackage{color}           
\usepackage{breakurl}        
\usepackage{textcomp}
\usepackage{hyperref}
\usepackage[utf8]{inputenc}

\newcommand{\aap}{    {\it Astron. Astrophys.}}

\newcommand{\apj}{    {\it Astrophys. J.}}
\newcommand{\apjl}{   {\it Astrophys. J. Lett.}}

\newcommand{\grl}{    {\it Geophys. Res. Lett.}}

\newcommand{\jgr}{    {\it J. Geophys. Res.}}

\newcommand{\solphys}{{\it Solar Phys.}}
 
\newcommand{\ssr}{    {\it Space Sci. Rev.}} 
\newcommand{\planss}{ {\it Planetary and Space Science}}

\chardef\us=`\_

\usepackage{rotating}

\begin{document}

\begin{article}
\begin{opening}

\title{Assessing the performance of EUHFORIA modeling the background solar wind}

\author[addressref={aff1,aff2},corref,email={juergen.hinterreiter@oeaw.ac.at}]{\inits{J.}\fnm{J\"urgen}~\lnm{Hinterreiter}}
\author[addressref=aff3,email={jasmina.magdalenic@oma.be}]{\inits{J.}\fnm{Jasmina}~\lnm{Magdalenic}}
\author[addressref=aff2,email={manuela.temmer@uni-graz.at}]{\inits{M.}\fnm{Manuela}~\lnm{Temmer}}
\author[addressref=aff4,email={Christine.Verbeke@kuleuven.be}]{\inits{C.}\fnm{Christine}~\lnm{Verbeke}}
\author[addressref={aff3,aff4},email={immanuel.jebaraj@oma.be}]{\inits{I. C.}\fnm{Immanuel Christopher}~\lnm{Jebaraj}}
\author[addressref={aff3,aff4},email={evangelia.samara@oma.be}]{\inits{E.}\fnm{Evangelia}~\lnm{Samara}}
\author[addressref={aff2,aff5},email={eleanna.asvestari@helsinki.fi}]{\inits{E.}\fnm{Eleanna}~\lnm{Asvestari}}
\author[addressref=aff4,email={stefaan.poedts@kuleuven.be}]{\inits{S.}\fnm{Stefaan}~\lnm{Poedts}}
\author[addressref=aff5,email={jens.pomoell@helsinki.fi}]{\inits{J.}\fnm{Jens}~\lnm{Pomoell}}
\author[addressref=aff5,email={emilia.kilpua@helsinki.fi}]{\inits{E. K. J.}\fnm{Emilia}~\lnm{Kilpua}}
\author[addressref=aff3,email={luciano.rodriguez@observatory.be}]{\inits{L.}\fnm{Luciano}~\lnm{Rodriguez}}
\author[addressref={aff3,aff4},email={camilla.scolini@kuleuven.be}]{\inits{C.}\fnm{Camilla}~\lnm{Scolini}}
\author[addressref=aff4,email={alexey.isavnin@kuleuven.be}]{\inits{A.}\fnm{Alexey}~\lnm{Isavnin}}

\address[id=aff1]{Space Research Institute, Austrian Academy of Sciences, Graz, Schmiedlstraße 6, 8042 Graz, Austria}
\address[id=aff2]{Institute of Physics, University of Graz, Universit\"atsplatz 5, 8010 Graz, Austria}
\address[id=aff3]{Solar–Terrestrial Centre of Excellence—SIDC, Royal Observatory of Belgium, 1180 Brussels, Belgium}
\address[id=aff4]{Centre for mathematical Plasma Astrophysics (CmPA), KU Leuven, 3001 Leuven, Belgium}
\address[id=aff5]{Department of Physics, University of Helsinki, P.O. Box 64, 00014 Helsinki, Finland}

\runningauthor{Hinterreiter et al.}
\runningtitle{EUHFORIA background solar wind modeling}

\begin{abstract}
In order to address the growing need for more accurate space weather predictions, a new model named EUHFORIA (EUropean Heliospheric FORecasting Information Asset) was recently developed \citep{pomoell18}. We present first results of the performance assessment for the solar wind modeling with EUHFORIA and identify possible limitations of its present setup. Using the basic EUHFORIA 1.0.4. model setup with the default input parameters, we modeled background solar wind (no coronal mass ejections) and compared the obtained results with ACE, in situ measurements. For the need of statistical study we developed a technique of combining daily EUHFORIA runs into continuous time series. The combined time series were derived for the years 2008 (low solar activity) and 2012 (high solar activity) from which in situ speed and density profiles were extracted. We find for the low activity phase a better match between model results and observations compared to the considered high activity time interval. The quality of the modeled solar wind parameters is found to be rather variable. Therefore, to better understand the obtained results we also qualitatively inspected characteristics of coronal holes, sources of the studied fast streams. We discuss how different characteristics of the coronal holes and input parameters to EUHFORIA influence the modeled fast solar wind, and suggest possibilities for the improvements of the model.
\end{abstract}

\end{opening}


\section{Introduction}
     \label{S-Introduction} 

The solar wind is a continuous flow of charged particles propagating radially outward from the hot corona of the Sun into interplanetary space. The speed measured at 1~AU heliocentric distance covers generally a range between 300 and 800~km~s$^{-1}$, consisting of slow solar wind and of high speed solar wind streams that have different characteristics and sources \citep[e.g.,][]{cranmer17,schwenn06}. 

The sources of the slow solar wind are closed magnetic field regions of coronal loops, active regions, coronal hole (CH) boundaries, but also streamers and pseudostreamers \citep{cranmer17}. On the other hand, fast solar wind emanates from open magnetic field regions, CHs, along which ionized atoms (mainly protons and alpha-particles) and electrons may easily escape to interplanetary space. CHs are localized regions of low density and low temperature in the solar corona that are generally slowly evolving and may persist for several solar rotations \citep{schwenn06}. However, where exactly within the CH the high speed component of the solar wind gets accelerated is not well understood and is subject of numerous studies.

High speed streams from CHs interact with the slower solar wind ahead causing compression regions that can lead to geomagnetic storms and the fast stream following the compression region with Alfvenic fluctuations can prolong substantially the recovery phase of the storm \citep[e.g.,][]{tsurutani1987}. It is well acknowledged that during the maximum phase of the solar cycle space weather is affected mostly by transient coronal mass ejections \citep[CME; e.g.,][]{webb2012}, however, during the declining and minimum activity phases high speed streams have significant impact \citep{tsurutani06,richardson2012,kilpua2017}. At all phases of solar cycle, high speed solar wind streams have also a paramount impact causing enhancements of Van Allen belt electron fluxes to relativistic electrons \citep[e.g.,][]{paulikas1979,jaynes2015,kilpua2015} and they strongly structure interplanetary space which is an important factor when studying and forecasting the propagation of CMEs. In general the morphology, area and location of CHs play a major role in the properties of the resulting compression region, duration and speed of the fast stream, and thus, its space weather impact level \citep[e.g.,][]{vrsnak07,garton18_2}. For example, statistical studies have shown that the equatorial parts of CHs are the main contributors to the fast solar wind streams measured at Earth \citep[see e.g.,][]{karachik11,hofmeister18} and that the speed of the solar wind at Earth increases with increasing CH area \citep[e.g.,][]{rotter12,nagakawa2019}. We note that with the evolution of a CH over time, also associated, the in-situ measured solar wind parameters can change \citep[e.g.,][]{heinemann18}. 



Models simulating the background solar wind are based on various methods, e.g., physics-based algorithms such as ENLIL \citep{odstrcil99} or MAS \citep{linker99} using synoptic photospheric magnetic field maps as input, empirical relations between observed areas of CHs and measured solar wind speeds at 1~AU \citep{vrsnak07,rotter12}, or simple persistence models using in-situ measurements shifted forward by variable time-spans depending on the spacecraft location \citep[e.g.][]{opitz09, owens13}. The performances of all the different solar wind models in comparison to actual measurements, reveal on average root-mean-square-errors of around 100--150~km~s$^{-1}$ in the wind speed and time shifts in the arrival of the peak speed of about $\pm$1~days and up to $\pm$3~days \citep[see e.g.,][]{owens08,macneice09,gressl14,reiss16,jian15,temmer18}. In general, model performances decrease with increased solar activity phases as CMEs frequently disturb the interplanetary space. Especially empirical solar wind models are not able to cope with those disturbances, but also for numerical models preconditioning is an important aspect which needs to be taken into account \citep{temmer17}. 

In order to address the growing need for more accurate space weather predictions, a new model named EUHFORIA (EUropean Heliospheric FORecasting Information Asset) was recently developed \citep{pomoell18}. In the following we present the first performance assessment of the solar wind model and identify possible caveats related to complex solar surface situations. 

\section{Solar wind modeling with EUHFORIA}
EUHFORIA is a physics-based simulation tool consisting of three essential parts: a coronal model, a heliospheric model and an eruption model. The main purpose of the coronal model is to provide realistic plasma conditions of the solar wind at the interface radius r~=~0.1~AU between the coronal and heliospheric model. The heliospheric model computes the time-dependent evolution of the plasma from the interface radius by numerically solving the MHD equations with the boundary conditions provided by the coronal model. For simulating transient events, CMEs are injected at the interface radius of the eruption model. Presently EUHFORIA shares similarities to the well-established solar wind/ICME model for the inner heliosphere, i.e. WSA-ENLIL \citep{Odstrcil2004}. An important feature of EUHFORIA is its flexibility. The three models, heliospheric, coronal and eruption one are fully autonomous and each part of EUHFORIA can be easily substituted with other models \cite[more details can be found in][]{pomoell18,scolini18}.  In contrast to ENLIL\footnote{We refer to ENLIL runs that can be performed at the Community Coordinated Modeling Center (CCMC) which can be found under \url{https://ccmc.gsfc.nasa.gov}} which gives the background solar wind parameters for a full Carrington rotation, EUHFORIA provides daily runs from hourly updated standard synoptic GONG magnetograms. In this way the central part of the magnetogram, used by EUHFORIA, is daily updated. For the purpose of the statistical studies and easier comparison with in situ observations we combine daily runs in order to obtain single time series (for a detailed description see Section 2.2).

In the present study we used EUHFORIA 1.0.4 version of the model, and we focus on the coronal and heliospheric model, in order to assess how well EUHFORIA simulates the background solar wind. For this study, we considered two phases of solar activity, one year during minimum in 2008 and another year during maximum in 2012. 

\subsection{The input parameters and setup of EUHFORIA}\label{Sec:DefaultParam}

\begin{figure}[htbp]
\centering
\includegraphics[width=\linewidth]{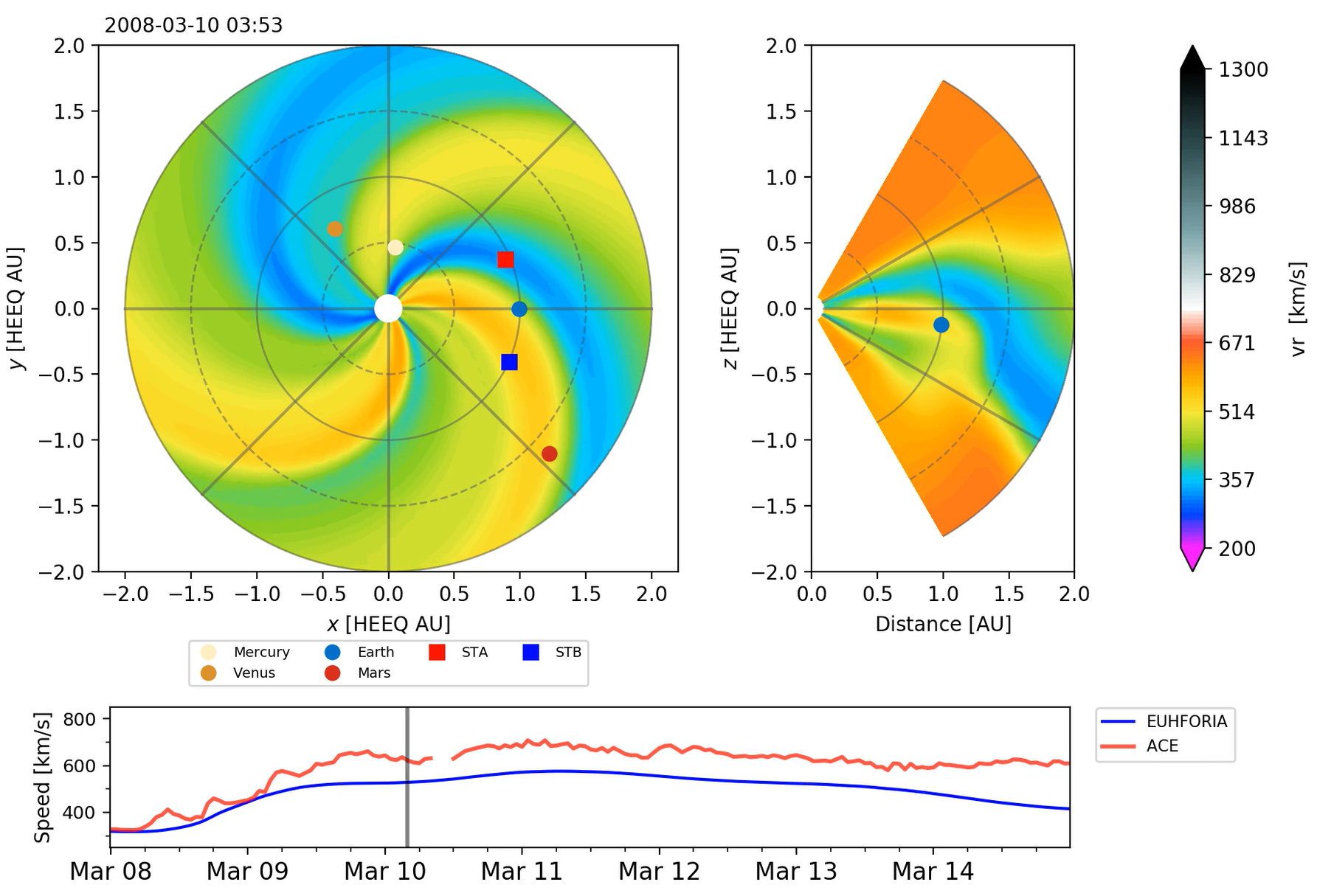}
\caption{Snapshot of the background solar wind radial speed modeled by EUHFORIA. The top left panel shows the MHD solution in the heliographic equatorial plane, and the right panel shows the meridional plane cut that includes the Earth (blue circle).
The lower panel shows comparison of the modeled and observed solar wind by EUHFORIA and ACE, respectively.}
\label{fig:EuhforiaSnapshot}
\end{figure}

As this is the first study of the solar wind modeling with EUHFORIA, we employed the so-called default setup that uses default values for the input parameters.
For the coronal part of the model, we use synoptic magnetograms from the Global Oscillation Network Group (GONG), and the potential field source surface (PFSS) model \citep{altschuler69} to simulate the magnetic field up to heights of 2.6~R$_{\odot}$ (so called source surface height). This is combined with the Schatten current sheet (SCS) model \citep{schatten69} starting from the height of 2.3~R$_{\odot}$ and that extends up to 0.1~AU. By overlapping the two models, a smoother transition between the lower coronal PFSS and upper coronal SCS model is obtained \citep[see][]{pomoell18, mcgregor08}. To determine the solar wind plasma parameters at the inner boundary of the heliospheric model we use the empirical Wang-Sheely-Arge model \citep{arge03} which is described below.

In EUHFORIA the solar wind speed depends on several parameters and the functional form of the empirical relation can be selected by the user. In this study we have employed the expression in the form:
\begin{equation}
    v(f,d)=v_0+\frac{v_1}{(1+f)^\alpha}\left[1-0.8{\rm exp}(-(d/w)^\beta\right]^3,
\label{Eq:EmpiricalForm}
\end{equation}
where $f$ and $d$ are the flux tube expansion factor and the great circular angular distance from the footpoint of each open field line to the nearest CH boundary, respectively. The parameters in Eq. \ref{Eq:EmpiricalForm} are set to $v_0$~=~240~km~s$^{-1}$, $v_1$~=~675~km~s$^{-1}$, $\alpha$~=~0.222, $\beta$~=~1.25 and $w$~=~0.02~rad. For a more detailed description see Eq. 2 in \cite{pomoell18}.
Since the original WSA relation is designed to provide the wind speed at Earth, and as the solar wind continues to accelerate beyond the inner boundary in the heliospheric MHD model, we have additionally subtracted 50~km~s$^{-1}$ to avoid a systematic overestimate of the wind speed. To compensate for the solar rotation, which is not included in the magnetic field model, we rotate the solar wind speed map at the inner boundary by 10°. We have also limited the minimum and the maximum solar wind speed at the inner boundary to 275 and 625~km~s$^{-1}$, respectively \citep[according to][]{mcgregor11}. 
In addition to the wind speed, the remaining MHD variables need to be determined. While the topology of the magnetic field is directly obtained from the SCS model, the magnitude of the solar wind magnetic field is set to be directly proportional to the speed. The plasma number density is given by
\begin{equation}
    n=n_{\rm fsw}(v_{\rm fsw}/v_r)^2,
\label{Eq:PlasmaNumberDensity}
\end{equation}
with the number density of the fast solar wind $n_{fsw}$~=~300~cm$^{-3}$ \citep[see e.g.,][]{bougeret84,venzmer18}, the fast solar wind speed $v_{fsw}$~=~675~km~s$^{-1}$ and $v_r$ coming from the empirical speed prescription. The maximum value $v_{\rm fsw}$~=~675~km~s$^{-1}$ is considered to be in the solar wind plasma with a magnetic field of 300~nT. For more details see Eq.~4 in \cite{pomoell18}.

Finally, we use a constant plasma thermal pressure of 3.3~nPa, at the inner boundary, that is in accordance with the fast solar wind temperature of about 0.8 MK. The angular resolution of the daily runs in this study was 4°, while 512 grid cells were chosen in the radial direction to cover the 0.1 to 2~AU domain.

An example of the background solar wind speed modeled by EUHFORIA, for the time interval of seven days in March 2008, is presented in Figure \ref{fig:EuhforiaSnapshot}. The two top panels (the heliographic equatorial and the meridional plane cuts plotted in the left and right panel, respectively) show that the Earth has entered a region of extended fast flow. The time of the snapshot is also marked by the black vertical line in the bottom panel which shows a comparison between the in situ observations and modeled solar wind speed. For this time period, we note a good match between the modeled solar wind by EUHFORIA and the in situ measurements (cf.\,bottom panel of Figure \ref{fig:EuhforiaSnapshot}).

\subsection{Combining individual runs and obtaining EUHFORIA time series}

\begin{figure}[htbp]
\centering
\includegraphics[width=\linewidth]{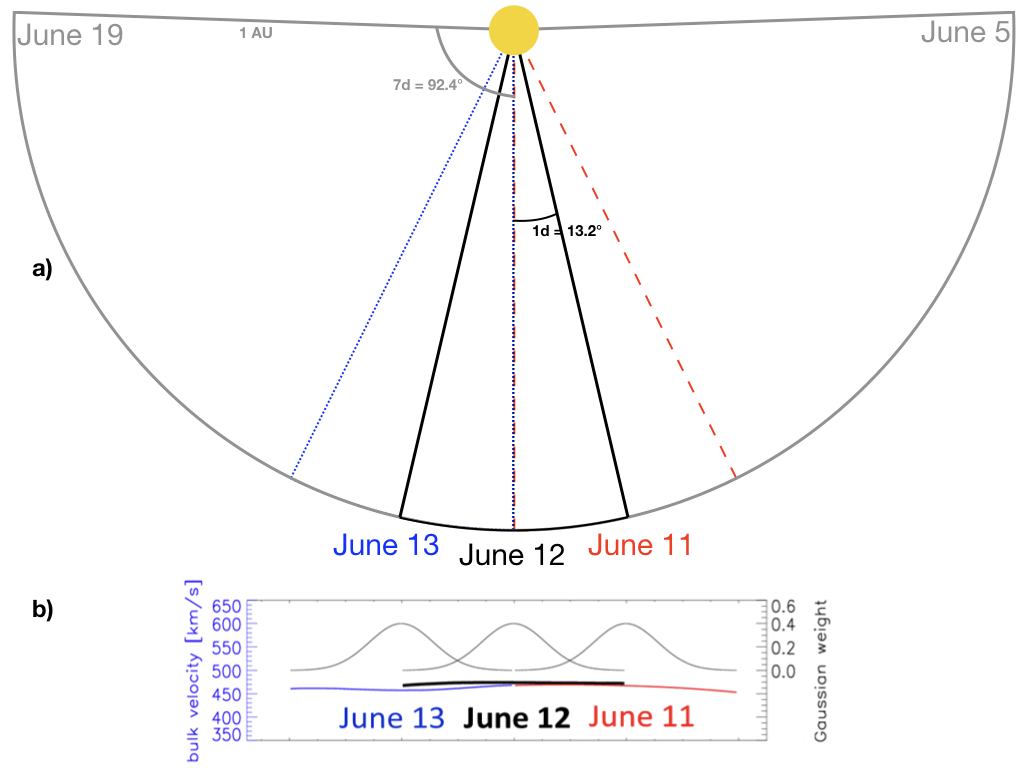}
\caption{Schematic representation of combining EUHFORIA model output for consecutive days. a) Different colors represent the selected range ($\pm$13.2° from the solar central region) for each day. Indicated in gray is the full range ($\pm$92.4°) provided by the model. b) Gaussian weight used for the model properties shown for three individual days.}
\label{fig:EuhforiaModel}
\end{figure}

For the systematic testing of the background solar wind, we used EUHFORIA daily runs, i.e., model outputs with default parameters, based on standard synoptic GONG magnetograms (the selected time was about 23:30 UT each day), for the complete years 2008 and 2012. We consider that each daily run, based on one magnetogram input, simulates the background solar wind at the heliocentric distance of 1~AU over a total time span of 14 days ($\pm$7 days) covering $\pm$92.4° in longitude (see gray slice in Figure \ref{fig:EuhforiaModel}) with a temporal resolution of 10~minutes. The central region of the Sun has the magnetic field information with the lowest projection effects, and is thus the most reliable part of the magnetogram. To combine the individual daily runs which overlap in time, we therefore developed a method containing information with highest weight on the central region of the Sun. The central region is defined as $\pm$1 day around the central meridian (0°) as given in the schematic drawing in Figure \ref{fig:EuhforiaModel}a. The weighting of each curve is done by a Gaussian distribution with the central part receiving the strongest weight (see Figure \ref{fig:EuhforiaModel}b).

\begin{figure}[htbp]
\centering
\includegraphics[width=\linewidth]{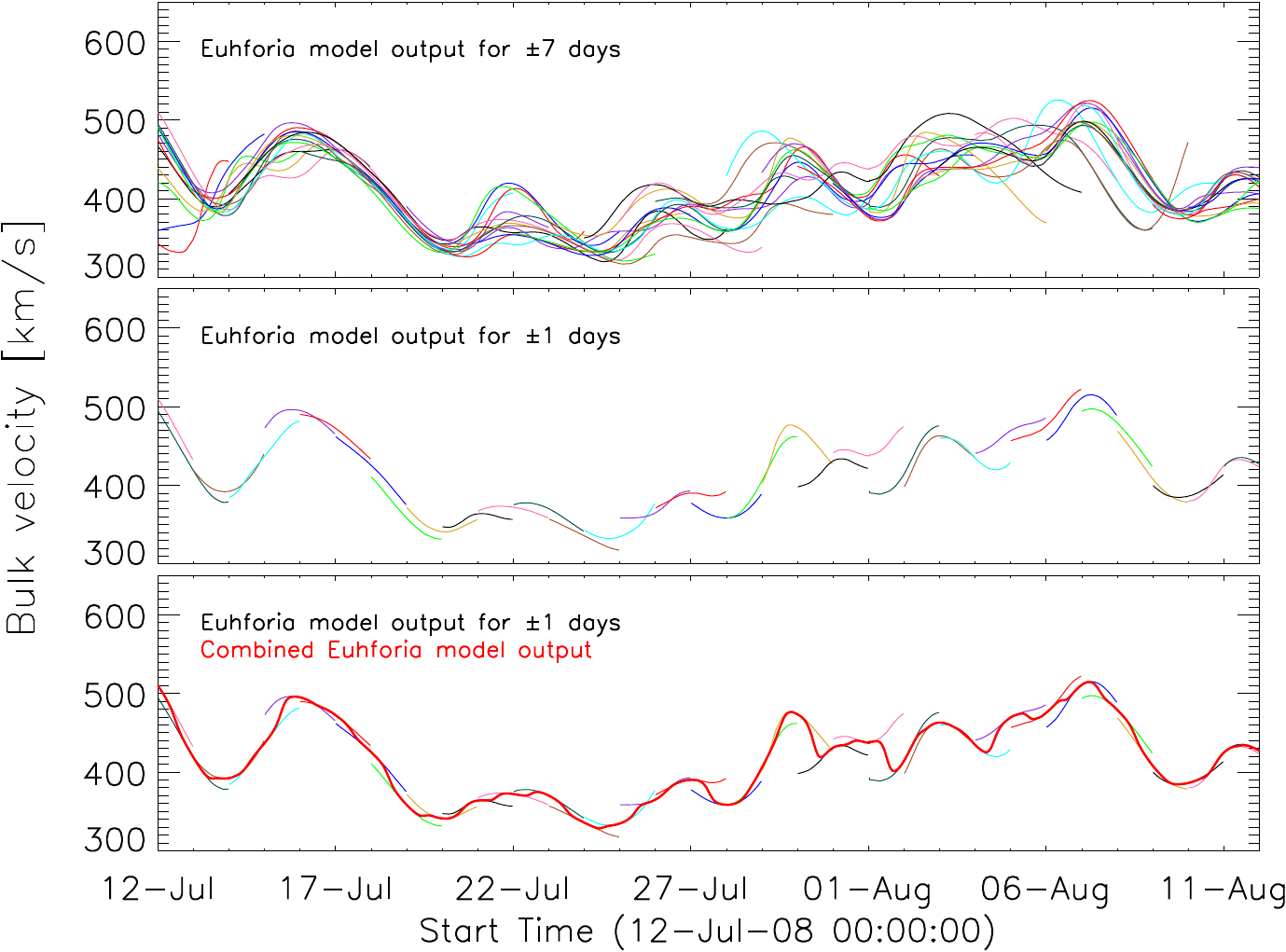}
\caption{Solar wind speed from July to August 2008. Top panel: Full EUHFORIA model output ($\pm$7 days). Middle panel: EUHFORIA model output limited to $\pm$1 day. Bottom panel: Model output (different colors for each daily run) and resulting time series (thick red).}
\label{fig:EuhforiaCurves}
\end{figure}

In Figure \ref{fig:EuhforiaCurves} we demonstrate how the method was applied. The top panel of Figure \ref{fig:EuhforiaCurves} shows the solar wind speed modeled by EUHFORIA for the full model output ($\pm$7 days). Different colors represent results from 32 daily runs. As can be seen, the simulated solar wind speeds for consecutive days may show significant offsets. In order to obtain a smooth time series we first limit the curves in time to $\pm$1 day (middle panel) and then combine them by using a Gaussian distribution (cf.\,Figure \ref{fig:EuhforiaModel}b). The obtained combined time series which is used for the analysis is given in the bottom panel of Figure \ref{fig:EuhforiaCurves} by the thick red curve. We also tested different limits of time ranges for the individual runs, e.g. $\pm$3~days, in order to check the quality of the method when combining individual runs. The resulting combined time series are rather similar and a bit more smoothed compared to using a time range limit of $\pm$1 day.

\begin{figure}[htbp]
\centering
\includegraphics[width=\linewidth]{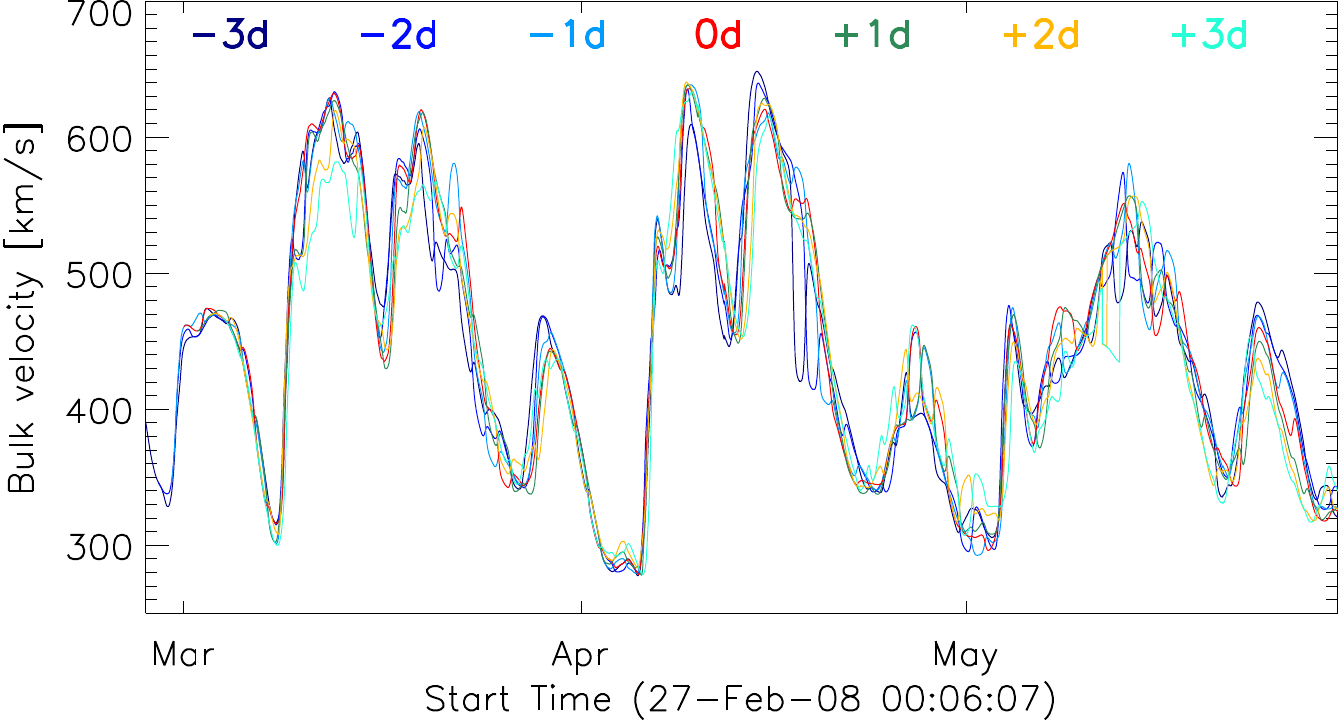}
\caption{Comparison of different shifts of the central region. The red curve (0d) represents the central region used for the individual runs. $-$3d indicates that the central region is shifted 3 days to the East while +3d indicates a shifting of the central region to the West.}
\label{fig:DifferentShifts}
\end{figure}

We evaluate how the combined time series for the modeled solar wind speed are affected when shifting the weighting to a region different than the central part of the Sun. With this, we take into account that comparing to the central region of the magnetogram the eastern or western region could influence more strongly the simulated solar wind. Figure \ref{fig:DifferentShifts} shows the results for the shifted weighting. One can observe clear differences between the combined time series, however, inspecting longer time ranges the general trend is retained.

\section{Comparison of in-situ observations and modeled solar wind}

\begin{figure}[htbp]
\centering
\includegraphics[width=\linewidth]{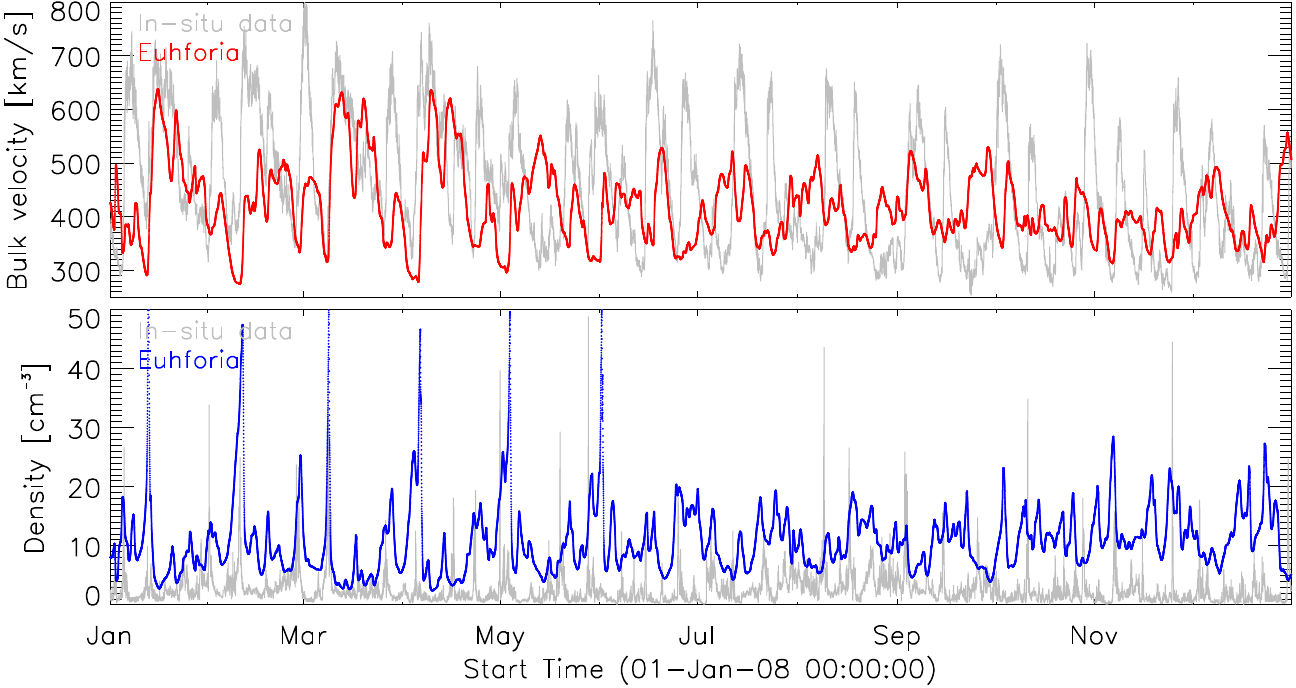}
\caption{EUHFORIA model output (red: velocity, blue: density) in comparison to in-situ measurements (gray) for 2008. Top panel: Solar wind bulk velocity. Bottom panel: Solar wind density.}
\label{fig:Euhforia2008}
\end{figure}

\begin{figure}[htbp]
\centering
\includegraphics[width=\linewidth]{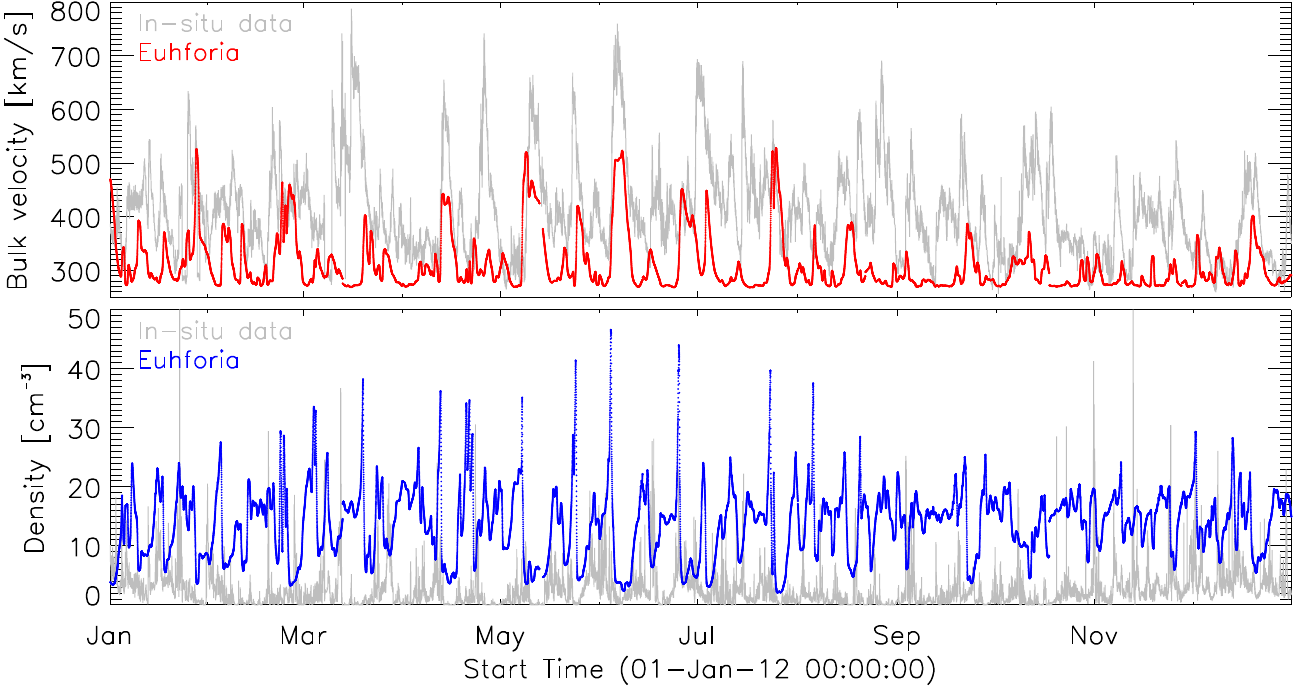}
\caption{Same as Figure \ref{fig:Euhforia2008} but for the year 2012.}
\label{fig:Euhforia2012}
\end{figure}

In order to asses the performance of the model we chose two intervals of different solar activity levels. At first, a quiet period during 2008 is considered, for which only three interplanetary coronal mass ejections (ICMEs), at the end of the year, were reported in the near-Earth solar wind according to the Richardson and Cane ICME list \citep[][see \url{http://www.srl.caltech.edu/ACE/ASC/DATA/level3/icmetable2.htm}]{richardson10}. This period can serve as benchmark time interval for the model performance as it almost optimally represents the background solar wind without significant transient perturbations. A second considered interval covers the year 2012, a period with rather high level of solar activity during which 35 ICMEs are reported (cf.\,Richardson and Cane ICME list). In order to evaluate how well EUHFORIA models the background solar wind, we compare the combined time series (see Section 2.2) with the in-situ measured plasma parameters speed and density as provided by the Solar Wind Electron, Proton and Alpha Monitor onboard the Advanced Composite Explorer \cite[SWEPAM/ACE,][]{mccomas98}. 

Figures \ref{fig:Euhforia2008} and \ref{fig:Euhforia2012} show the results obtained for the years 2008 and 2012. The gray curves represent observed values by ACE, while red and blue curves represent modeled values of the solar wind speed and density, respectively. The presented statistics of the background solar wind modeled with EUHFORIA shows on average lower values of the modeled solar wind speed than the in-situ measured velocity. On the other hand the modeled solar wind density is considerably higher than the observed one. In the present setup of EUHFORIA these two solar wind plasma parameters are coupled (cf.\,Eq.~\ref{Eq:PlasmaNumberDensity}), and improved modeling of the solar wind speed will also result in a better modeled solar wind density. We also noticed that the correlation between modeled and observed values is significantly better in the first half of year 2008 (Figure \ref{fig:Euhforia2008}). In the second half of year 2008, the maximum speeds for the fast solar wind speed are not well modeled by EUHFORIA, and also the minimum values are significantly different, i.e., larger than the observed ones. For the year 2012 the discrepancies between the modeled values and observations are more pronounced. Nevertheless, periods of lower wind speeds during 2012 are rather well reproduced, which might be simply a consequence of a very low wind speed in general obtained for this year.

The in-situ solar wind speed, for both studied years, was also compared to the individual daily runs in order to assess the probability of artificially enhanced or reduced fast wind flows due to combining of the daily runs (Section 2.3). In the two studied years we found only one case of the fast solar wind which was observed in the majority of the daily runs but not in the combined time series (around 22 August 2012). The opposite cases, where the combined time series show significant increase of the solar wind speed that was not modeled in the majority of the relevant daily runs, were not found. 

As a consequence of the, on average, underestimated solar wind speed modeled by EUHFORIA, fast flows arrive with a systematic delay in time. The amount of delay depends on the difference between the modeled and observed wind speed. For example, the fast solar wind with average speed of 600~km~s$^{-1}$ will need about 2.9~days to arrive to the Earth, while those of about 500~km~s$^{-1}$ will need about 3.5~days. In this case the induced latency of modeled solar wind will be about 14~hours. We observe the influence of this effect particularly strong in the second half of the year 2008 (Figure \ref{fig:Euhforia2008}).

\subsection{Evaluation of modeling results}

In order to evaluate the EUHFORIA model performance we present a hit-miss statistics using two different methods for comparing measured and modeled results. We also compare the minimum and maximum phase of the results and give initial results on the effects of different input parameters for the model. In this analysis we focus only on the solar wind velocity. 

\subsubsection{Hit-miss statistics by automatic peak-peak matching method}
To evaluate the model performance, we calculate continuous variables (e.g., Root-mean-squared error RMSE) and apply an event-based approach for detecting the maxima (peak finding algorithm) in the solar wind observations. For the event-based approach we used an automatic peak finding algorithm. To be defined as a peak, certain properties \cite[minimum speed~=~400~km~s$^{-1}$, minimum gradient~=~60~km~s$^{-1}$, for further details see][]{reiss16} 
have to be fulfilled. A hit is found, if the modeled peak appears within a time window of $\pm$2~days around the measured peak, and a miss if the modeled peak is out of this time window. If the peak is found in the combined time series of EUHFORIA and not in observations we consider to have a false alarm.

\begin{sidewaysfigure}
\centering
\includegraphics[width=\linewidth]{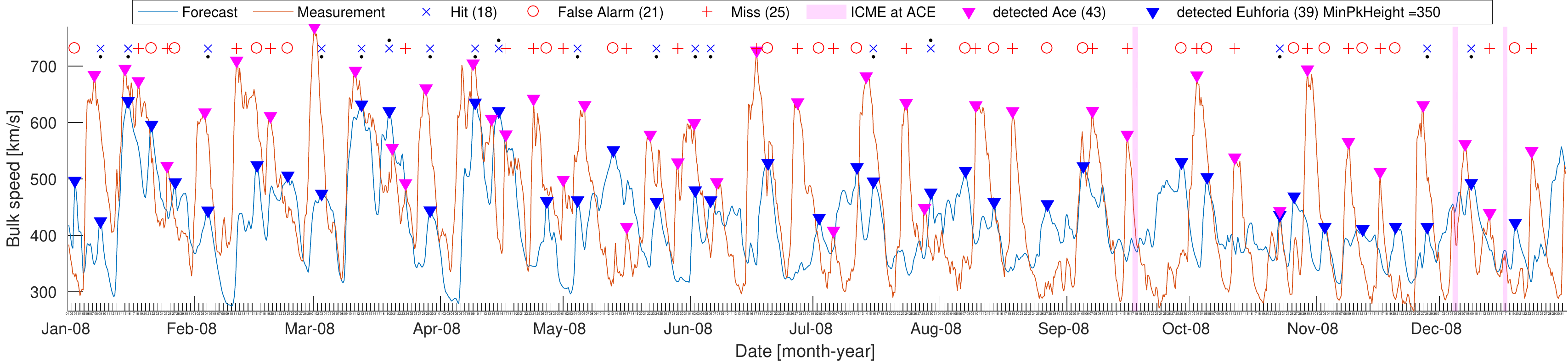}
\includegraphics[width=\linewidth]{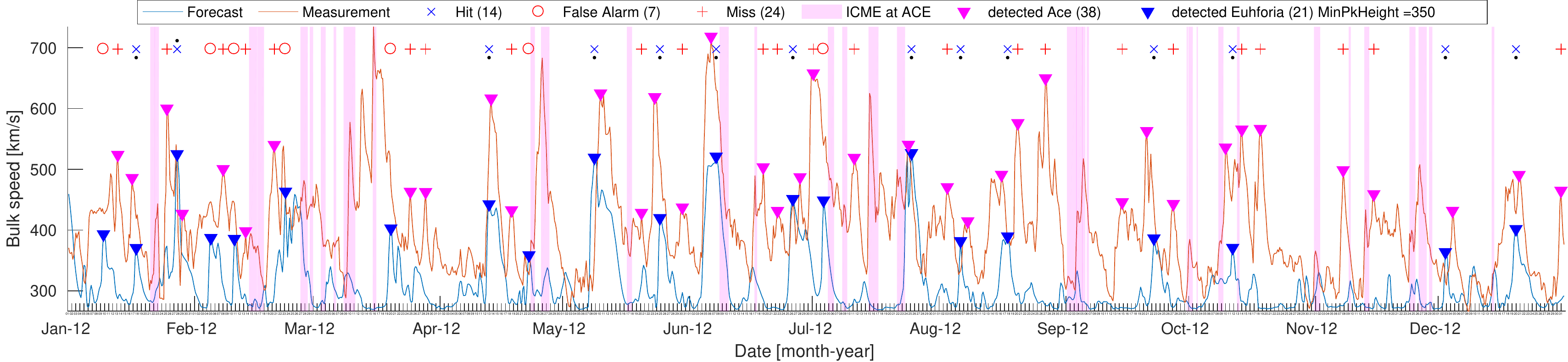}
\caption{EUHFORIA modeled solar wind bulk velocity (blue) in comparison to in-situ measurements (orange) for 2008 (top) and 2012 (bottom) using a peak finding algorithm. The red vertical bars indicate times of CME occurrences according to \cite{richardson10}. }
\label{fig:PeaksEuhforia}
\end{sidewaysfigure}

Since the study encompasses also the year 2012 with the high level of solar activity, it was necessary to isolate intervals with possible ICMEs in the in-situ observations. The vertical pink lines in Figure \ref{fig:PeaksEuhforia} indicate the times of CME occurrences according to the CME list \citep{richardson10}. We note that for 2008 only three ICMEs were reported while for 2012 there are 35 reported events. 

For both years under study we obatin a similar result for the RMSE which is about 125~km~s$^{-1}$. As can be seen from Figure \ref{fig:PeaksEuhforia}, in 2008 (top panel), 39 solar wind peaks are detected in the EUHFORIA combined time series and 43 in the in-situ data. Applying the automatic peak finding algorithm method, we obtain 18 hits, 21 false alarms and 25 misses. In 2012 (bottom panel in Figure \ref{fig:PeaksEuhforia}), the EUHFORIA combined time series shows 21 peaks and 38 are detected in the in-situ observations. This corresponds to 14 hits, 7 false alarms and 24 misses. As this is a rather poor result we inspect the solar wind profiles (observed and modeled) in more detail and investigate the reason of the poor performance. 

\begin{figure}[htbp]
\centering
\includegraphics[width=0.95\linewidth]{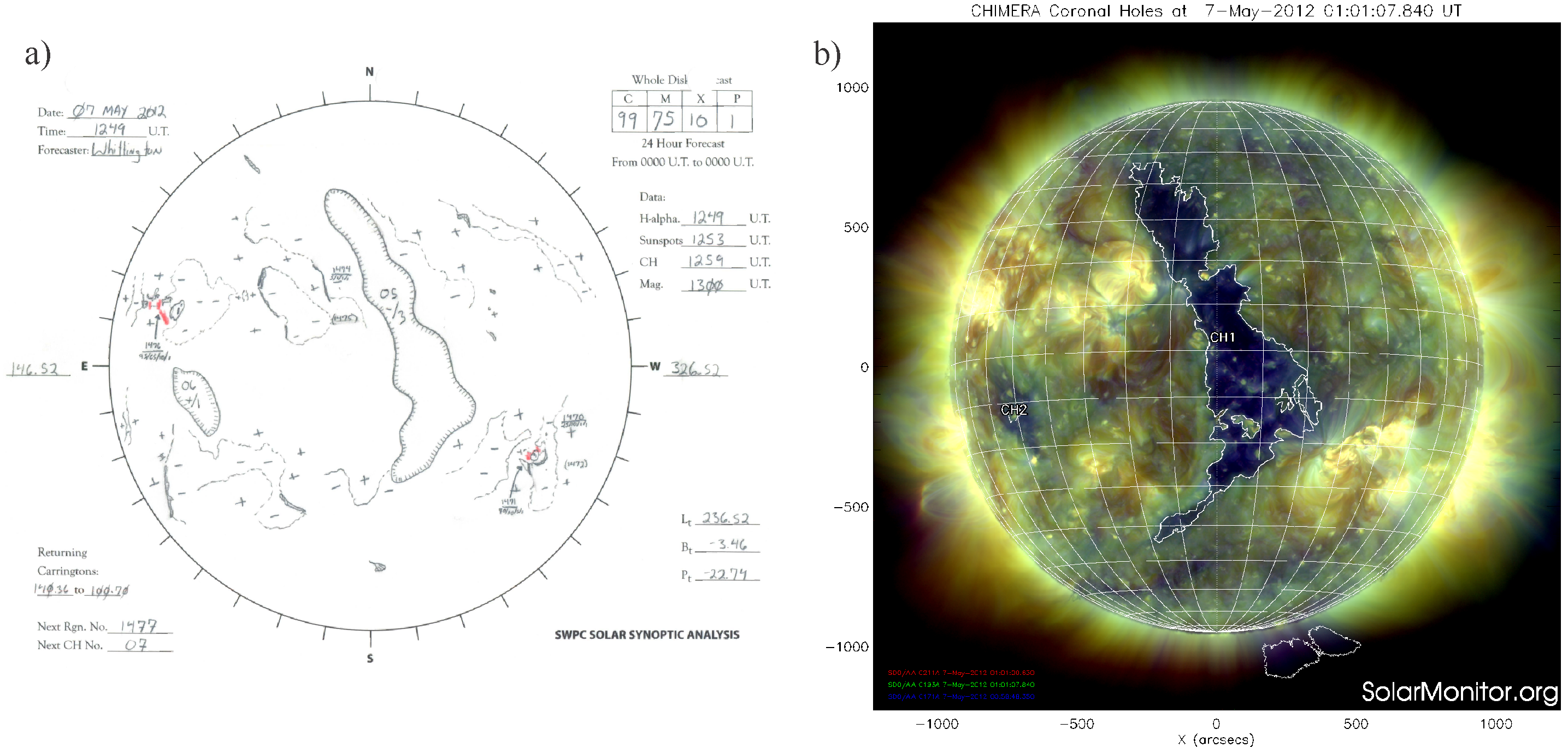}
\caption{a) Drawing of the solar surface features for 2012 May 7 provided by \href{https://www.ngdc.noaa.gov/stp/space-weather/solar-data/solar-imagery/composites/full-sun-drawings/boulder/}{NOAA}.
The CH is identified using EUV imagery from spacecraft while the polarity of the CH is obtained from magnetograms. b) Detection of the CH on the same day, by the CHIMERA tool (based on three wavelengths 211, 193, 171\AA). The image was obtained from \href{https://www.solarmonitor.org/}{Solar Monitor}.}
\label{fig:Drawing_CHIMERA}
\end{figure}

\subsubsection{Hit-miss statistics by manual peak-peak matching method}

The in-situ observations frequently show several subsequent local maxima of the solar wind speed associated with a single fast flow generally originating from a large and extended, in latitude or in longitude or both, CH. In such a case the automatic peak finding algorithm finds several peaks and it is not possible to make a one-to-one identification with the usually smooth increase of the solar wind speed modeled by EUHFORIA. In order to better understand such long lasting flows and to unambiguously relate modeled and observed velocity peaks with each other, we checked the development of the CHs on the Sun two days before and three days after the CH started its transition across the central meridian (see Figure \ref{fig:Drawing_CHIMERA}). 
For this purpose we analyzed automatic CH areas detected by the CHIMERA software \citep{garton18_1} and CH drawings (see Figure \ref{fig:Drawing_CHIMERA}).

As for the automatic method, the intervals corresponding to ICME arrivals, reported in a list by \cite{richardson10} and observed in-situ, were excluded from the evaluation. In addition, peaks in the in-situ measured solar wind speed that could not be related to CHs were also excluded from the statistical study. We considered observed and modeled solar wind peaks to be associated, i.e. a hit, if the increase started more or less simultaneously and the peak was achieved within 2 days after the peak as modeled by EUHFORIA. When the modeled solar wind increase did not have the counterpart in the in-situ observations we considered to have a false alarm, and when the observed fast flow was not reproduced by EUHFORIA we consider to have a miss.  

The manual identification of the CHs and associated fast flows shows 17 hits, 12 misses and 6 false alarms for 2008 and 13 hits, 18 misses and 0 false alarms for 2012. We note that these results reveal a significantly smaller number of false alarms and misses in comparison to the automatic method. This indicates that the CH development and its shape has strong influence on the fast solar wind speed profile measured at 1~AU.

\subsubsection{Solar cycle dependence}

In Figures \ref{fig:Euhforia2008} and \ref{fig:Euhforia2012} can be seen that the solar wind modeled by EUHFORIA matches much better for the interval of the minimum solar activity in 2008. This may have several reasons. During the low level of solar activity the magnetic field, the main input for the PFSS extrapolation in EUHFORIA, changes less dynamically than during the high level of solar activity, which can result in a more reliable modeling of the solar wind flow. Also, the interplanetary measurements are not disturbed by transient events which are much less frequent compared to solar maximum activity, and the solar wind flow is more persistent \citep{owens13, temmer18}. 

Figure \ref{fig:Euhforia2008}a shows for 2008 on average rather good model results of the minimum and maximum solar wind speed, and the majority of fast flows associated with equatorial CHs is well reproduced. However, we also found an exception where the in-situ observations show a recurrent fast flow (10 rotations) associated with a well-defined equatorial CH which was modeled by EUHFORIA only at the beginning of the year 2008. We believe that modeling of the solar wind originating from this particular CH is highly influenced by the CH characteristics and development in location, size and shape.

During the high level of solar activity the magnetic field is very complex and it is known that the amount of low latitude open flux may be significantly underestimated by the PFSS model \citep[e.g.][]{macneice11}. Underestimating the open flux leads to significantly lower solar wind speeds modeled by EUHFORIA. This effect is very strongly pronounced in 2012 (Figure \ref{fig:Euhforia2012}a). We also note for 2012 the existence of a large number of low latitude CHs surrounded by active regions which possibly also influences the model performance by strongly deviating the magnetic topology from being potential.

\subsection{Identified limitations of the basic setup of EUHFORIA}

During testing of the modeled background solar wind we identified some limitations of the present version of EUHFORIA which influences its performance. Herein we identify some of the limitations of the basic setup of the EUHFORIA 1.0.4. and a more detailed analysis will be presented in the follow up paper by Samara et al.\,(2019; in preparation).

\begin{figure}[htbp]
\centering
\includegraphics[width=\linewidth]{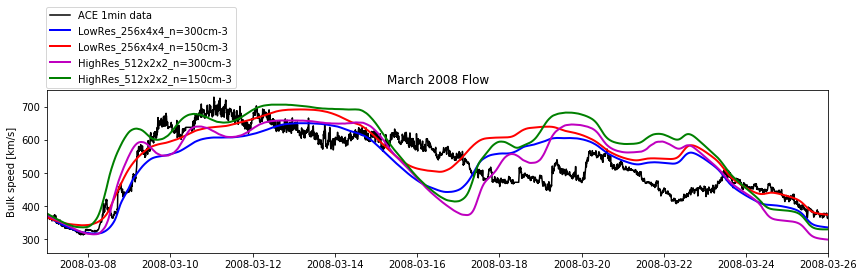}
\caption{Comparison of model runs with different settings. High/default density: 300 $cm^{-3}$, low/default resolution: 4° in longitude and latitude with 256 radial cells (256x30x90). Low density: 150 $cm^{-3}$, high resolution: 2° in longitude and latitude with 1024 radial cells (1024x60x180).}
\label{fig:ComparisonMarch2008}
\end{figure}

\subsubsection{The default input parameters of EUHFORIA}

In order to set up benchmarks for the solar wind modeling with EUHFORIA we need to understand how different input parameters influence the modeled solar wind. 
Figure \ref{fig:ComparisonMarch2008} shows the EUHFORIA model results for the time interval of several days in March 2008 using different input parameters. We vary the resolution of the heliospheric model and the input density of the fast solar wind at the inner boundary compared to the default setting \citep[Section 2.1 herein and Section 2.1.2. in][]{pomoell18}. We find that a decrease of the solar wind density by 50\% (initial value is 300 cm$^{-3}$ at 21.5~R$_{\odot}$) induces an increase of the modeled solar wind speed from several percent up to 15\% (absolute value depends on the part of the flow which is considered). Figure \ref{fig:ComparisonMarch2008} also shows a comparison of the default, low resolution runs (angular and radial resolution of 4° and 256 cells, respectively) and the intermediate resolution runs (2° and 512 cells, respectively). The higher resolution runs result in an increased solar wind speed (up to about 20\%) and in an earlier arrival time of the high speed stream at 1~AU (up to several hours). If we compare the two extreme cases, the default EUHFORIA runs i.e., low resolution and high density, and the intermediate resolution and low density runs, we find a shift of the arrival time of the fast flow of about $-$12 hours, and a significant increase of the solar wind speed (from about 6\% to more than 40\%, depending on which part of the fast flow is considered). 
The obtained results indicate that the quality of the modeled fast solar wind varies a lot depending on the input parameters to the model. We note that when more than one parameter is modified the solar wind speed changes in a non-linear manner and that the changes strongly depend on the considered flow. This brings forward the need for a detailed ensemble parameter study which will provide a well-defined benchmark for the solar wind modeling with EUHFORIA (Samara et al., in preparation).

\subsubsection{Open flux and the source surface heights}

Comparing CH sizes extracted from EUV observations, and modeled open flux areas (i.e., CH areas) by PFSS using GONG synoptic magnetograms, shows that on average CHs are underestimated in the model. It is found that the amount of modeled open flux is lower than actually observed, as well as open flux areas show up smaller in angular width \citep{asvestari19}. Failure in reliably modeling open magnetic flux has consequences for a proper solar wind modeling, in particular for the fast solar wind flow originating from CH areas. This will result not only in an underestimation of the solar wind speed but also might cause the fast flow to be too narrow, hence, may completely miss the Earth (Section 3.2.3). In a systematic testing it was shown that changing the source surface height (one of the default input parameters to EUHFORIA) significantly influences the modeled open flux and can result even in a shift of the position of the considered CH \citep{asvestari19}.

\begin{figure}[htbp]
\centering
\includegraphics[width=\linewidth]{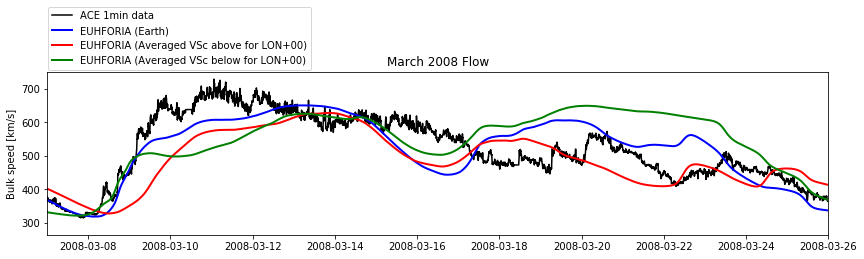}
\caption{Combined EUHFORIA results in comparison to in-situ data (gray) for the same interval as in Figure \ref{fig:ComparisonMarch2008}.
Earth (blue curve) represents the EUHFORIA output for the Earth location. The red curve is an average of combined EUHFORIA results for virtual spacecraft (+4°, +8°, +12°) above the Earth and the green curve shows the averaged results for virtual spacecraft ($-$4°, $-$8°, $-$12°) below the Earth.}
\label{fig:ComparisonVSC}
\end{figure}

\subsubsection{Dependence on shape and location of CHs}

While manually associating the observed and modeled solar wind flows (Section 3.1.2.), we recognized that the EUHFORIA performance is closely related also to the size, shape and location of the CHs, sources of the fast flows. The qualitative study of the CH characteristics and the quality of the modeled fast solar wind (Section \ref{Sec:DefaultParam}) shows that for circular and equatorial CHs occurring during the low level of solar activity, EUHFORIA models well the associated fast flows. However, fast flows associated with narrow CHs elongated in longitude, are rarely reproduced well by EUHFORIA. In the case of the narrow CHs elongated in latitude, the modeled solar wind is mostly underestimated, hence, leading to a late arrival at the Earth. And when the solar wind is originating from the low/high latitude CHs (greater than $\pm$30°) and/or the extensions of the polar CHs, it will be rarely reproduced correctly by EUHFORIA. We also noticed that fast flows associated with patchy CHs, irrespective of their latitudes and longitudes, are poorly reproduced or not reproduced at all by EUHFORIA.

Further on, the fast flows originating from low latitude CHs might pass 'below' or 'above' the Earth (when the associated CHs are situated at the southern or northern solar hemisphere, respectively) and they will not be observed in the EUHFORIA time series output at the Earth \citep[see also][]{hofmeister18}. In order to check this hypothesis we have implemented virtual spacecraft around the Earth (separated by 4° ranging from $-$12° to +12° in latitude where 0° indicates Earth position) and compared the modeled time series for all these spacecraft. To amplify the effect, the values of time series at +4°, +8°, +12° above the Earth and $-$4°, $-$8°, $-$12° below the Earth were averaged and compared to in-situ data (see Figure \ref{fig:ComparisonVSC}). We note that the fast flow, starting on March 09, 2008 seem to be reproduced well by EUHFORIA, by all three time series, i.e. above the Earth, at the Earth and below the Earth. This gives indications on the 3D extent of the fast flow that directly impacted the Earth, which is also visible in Figure~ \ref{fig:EuhforiaSnapshot} right top panel. The solar wind observed starting from March 19 (Figure \ref{fig:ComparisonVSC}) originates from rather large low latitude extensions of the southern polar CH. EUHFORIA models at the Earth a somewhat faster solar wind then observed by ACE (blue curve), and significantly faster solar wind passing below the Earth (green curve). In this case the fast flow only glanced the Earth while the main part of the fast solar wind passed below the Earth. Studies of the 3D extent of the fast flows, using the virtual spacecrafts, is among the main ongoing efforts for improving our knowledge on the solar wind and solar wind modeling with EUHFORIA (Samara et al., in preparation).

\section{Summary and Conclusions}

In this paper we present the first results of the solar wind modeling with the new European model EUHFORIA. For the statistical study we employed the so called basic setup of EUHFORIA 1.0.4. using default input parameters (Section \ref{Sec:DefaultParam}). EUHFORIA currently provides daily modeled results using synoptic GONG magnetograms. In order to obtain a continuous time series of the background solar wind parameters, the model outputs from consecutive days have to be combined. We developed a method to derive such a continuous profile from individual runs taking only the central part of the individual curves and combining them using a Gaussian weighting (Section 2.2).

We test the quality of the performance of EUHFORIA in solar wind modeling by selecting two years of different solar activity levels, i.e. 2008 and 2012. The analysis was focused on the comparison of the modeled solar wind for the two most important solar wind plasma parameters, i.e. bulk speed and proton density, and ACE observations (Figures \ref{fig:Euhforia2008} and \ref{fig:Euhforia2012}). As a general trend we notice an underestimation of the modeled solar wind speed and an overestimation of the modeled density, in comparison with in situ observations by ACE. The solar wind modeled by EUHFORIA matches better for the interval of the minimum solar activity in 2008, then for the year 2012 when the level of solar activity was high. We conclude that this result is mostly originating from the better performance of the PFSS model (main part of the EUHFORIA's coronal model) during low level of the solar activity. 

For defining the association between modeled and observed fast flows we applied an automatic peak finding algorithm (Section 3.1.1). Using this algorithm we obtain 18 hits, 21 false alarms and 25 misses for 2008 and 14 hits, 7 false alarms and 24 misses for 2012. As a consequence of the frequently underestimated solar wind speed modeled by EUHFORIA, arises the uncertainty in the modeled arrival time of fast streams. Moreover, depending on the CH shape and location on the Sun, fast single flows may show multiple wind speed maxima which restricts the automatic peak finding algorithm in finding the correctly matching pairs. By visual inspection (Section 3.1.2) we took into account all these characteristics and assign more reliably the modeled and measured solar wind flow pairs, and obtained better statistics of 7 hits, 6 false alarms, and 12 misses for 2008 and 13 hits, 0 false alarms, and 18 misses for 2012.

Our statistics show that the quality of the modeled fast solar wind, obtained using the basic setup of EUHFORIA and the default input parameters, can be very variable. In the current study we identified some of the limitations of this setup. E.g., a higher angular resolution from 4° to 2° can result in an increase of the solar wind speed by up to 20\% and with that causes an earlier arrival of the fast solar wind up to several hours. Additionally, as expected high resolution runs show significantly more structures in the solar wind in comparison to the low resolution ones. We also tested how the decrease of the fast solar wind density from 300 $cm^{-3}$ to 150 $cm^{-3}$ influences the modeled solar wind and found that in the case of the lower input density EUHFORIA will model earlier arrival and larger amplitudes of the fast solar wind (Section 3.2.1.). When combined, even only these two factors can lead to substantial errors in predictions. Detailed analyzes on such limiting factors are presented in follow-up studies by \cite{asvestari19} and Samara et al.\,(2019; in preparation). 

The visual inspection of the CHs associated to the fast flows indicates, that the shape and the location of the CHs play an essential role in the model performance (Section 3.2.3). We found that patchy, elongated and narrow CHs are not well simulated by EUHFORIA's coronal model (i.e., PFSS misses open flux), which results in a poor model performance. We also found that the high latitude ($>$ 30°) CHs, often extensions of polar CHs, may be responsible for EUHFORIA modeling the fast flow passing above or below the Earth (in a case of CHs on the northern and southern solar hemisphere, respectively). Therefore, it is very important to have EUHFORIA set up with included virtual spacecraft for all the future studies of the solar wind modeling by EUHFORIA. This will allow us to estimate the 3D extend of the fast flows and to understand if the fast flow just missed the Earth, passing below or above it (Section 3.2.3).

In the herein presented studies we identified some of the limitations of the present version of EUHFORIA 1.0.4. which influences its performance, in particular during the high level of solar activity. We found that the dynamic behaviour of the CHs, together with the complex coronal magnetic field has a major role in the generation and propagation of the fast solar wind. Due to the complexity of the solar atmosphere modeling of the fast solar wind is a very demanding task. Herein we presented first attempts to model background solar wind with EUHFORIA, identified some of the limitations of the present setup of the model and presented first example of the parameter studies. The presented results bring forward the need for a detailed ensemble parameter study which will provide a clear benchmark for the solar wind modeling with EUHFORIA, but which goes beyond the scope of this paper. The parameter studies, which are presently ongoing in the framework of the CCSOM project (\url{http://www.sidc.be/ccsom/}), will help us not only to improve modeling of the solar wind with EUHFORIA but also to improve EUHFORIA itself.

\begin{acks}
 J.H. acknowledges the support by the Austrian Science Fund (FWF): P 31265-N27. M.T. acknowledges the support by the FFG/ASAP Program under grant No. 859729 (SWAMI). E.A. would like to acknowledge the financial support by the Finnish Academy of Science and Letters via the Postdoc Pool funding. C.S. was funded by the Research Foundation – Flanders (FWO) SB PhD fellowship no. 1S42817N. E.A. acknowledges the support by the Finnish Academy of Science and Letters via a Postdoc Pool grant.
\end{acks}





\bibliographystyle{spr-mp-sola}

\IfFileExists{\jobname.bbl}{} {\typeout{}
\typeout{****************************************************}
\typeout{****************************************************}
\typeout{** Please run "bibtex \jobname" to obtain} \typeout{**
the bibliography and then re-run LaTeX} \typeout{** twice to fix
the references !}
\typeout{****************************************************}
\typeout{****************************************************}
\typeout{}}

\end{article} 

\end{document}